\documentclass[aps,onecolumn,showpacs]{revtex4}
\usepackage{graphicx}
\usepackage{epsfig}
\usepackage{amssymb}
\usepackage{graphicx}
\usepackage{amsfonts}
\usepackage{latexsym}
\usepackage{amsmath}
\usepackage{subfigure}
\usepackage{mathrsfs}

\newcommand{\tpsi}
{\widehat\Psi}

\newcommand{\bt}{(\beta)}

\def\pre#1#2#3{Phys. Rev. E  {#1} #2 #3}

\begin{document}
\title{Generalized Christoffel-Darboux formula for
skew-orthogonal polynomials and random matrix theory}
\author{Saugata Ghosh}\email{sghosh@ictp.it}
\affiliation{The Abdus Salam ICTP, Strada Costiera 11, 34100,
Trieste, Italy.}
\date{\today}
\begin{abstract}
We obtain a generalized Christoffel-Darboux (GCD) formula for
skew-orthogonal polynomials. Using this, we present an alternative
derivation of the level density and two-point function for
Gaussian orthogonal ensembles and Gaussian symplectic ensembles of
random matrices.
\end{abstract}
\maketitle

Random matrices have found applications in different branches of
physics  mainly due to the `universality' in their correlation
function under certain scaling limits. In this context, although
unitary ensembles of random matrices, corresponding to systems
with broken time-reversal symmetry (for example, a mesoscopic
conductor in the presence of a magnetic field) have been
extensively studied \cite{mehta,eynard1}, much less is known about
Orthogonal Ensembles (OE) and Symplectic Ensembles (SE) of random
matrices.

In his phenomenal paper \cite{dyson1} Dyson stressed the
importance of a good understanding of the skew-orthogonal
polynomials (SOP) to study the OE and SE of random matrices. Some
progress has been made in this regard \cite{mehta,ghosh,
ghoshpandey} to develop the theory of SOP. Alternative approaches
were taken by various authors \cite{stoj1,
stoj1a,stoj2,af,f,tw1,tw3,tw4} to study these ensembles. For
example,  Deift and Gioev \cite{deift1,deift2} have recently used
the Widom's representation \cite{w} to prove Universality for a
wide class of polynomial potentials of the OE and SE of random
matrices. In all these work, the authors have used the well known
properties of orthogonal polynomials to study these ensembles.
Here we use the more elegant SOP, evolving naturally from OE and
SE of random matrices.

In this paper we obtain recursion relations  for SOP. Using this
we derive the GCD formula. We use them to obtain the level density
and two-point correlation function for Gaussian orthogonal
ensembles and Gaussian symplectic ensembles of random matrices.

We consider  ensembles of  $2N$
dimensional  matrices
$H$  with probability distribution
\begin{equation}
P_{\beta,N}(H)dH=\frac{1}{{\cal Z}_{\beta N}}\exp[-[{\rm
Tr}V(H)]]dH,
\end{equation}
where the parameter $\beta=1$ and $4$ corresponds to $H$  real
symmetric or quaternion real self dual.(Note that we have
considered the OE of even dimension. The odd dimension can be
easily generalized.) $dH$ is the standard Haar measure. ${\cal
Z}_{\beta N}$ is the so called `partition function', and is
proportional to the product of skew-normalization constants
\cite{mehta}:
\begin{equation}
{\cal Z}_{\beta N}= \int\exp[-[{\rm Tr}V(H)]]dH
=N!\prod_{j=0}^{n_F-1}g_{j}^{\bt}.
\end{equation}
Here $n_F$ is called the `Fermi level' by analogy with a system of
fermions. In our case, it takes the value $n_F=2N$ for $\beta=1$
and $4$ respectively. It is assumed that most of the physical
quantities (like level density and correlation functions) are
related to properties of $g_n^{\bt}$ around the `vicinity'
\cite{eynard} of the Fermi level. In this paper, we give rigorous
justification for such a claim.


To study different correlations among the eigenvalues of random
matrices, we need to study certain kernel functions
\cite{dyson1,mehta}. For example, the two-point correlation
function for $\beta=1$ and $4$ can be expressed in terms of the
$2\times 2$ matrix

\begin{eqnarray}
{\sigma}^{(\beta)}_2(x,y) &=& \left(\begin{array}{cc}
S^{(\beta)}_{2N}(x,y) & D^{(\beta)}_{2N}(x,y)     \\
I^{(\beta)}_{2N}(x,y) & S^{(\beta)}_{2N}(y,x)         \\
\end{array}\right),
\end{eqnarray}
while the level-density
\begin{equation}
\rho^{(\beta)}(x)=S^{(\beta)}_{2N}(x,x).
\end{equation}
Here, the kernel functions are defined as
\begin{eqnarray}
\label{kernel} \nonumber S^{(\beta)}_{2N}(x,y) &=&
-{{\widehat{\Phi}}^{(\beta)}}(x)\prod_{2N} \Psi^{(\beta)}(y),\\
&=&{{\widehat{\Psi}}^{(\beta)}}(y)\prod_{2N}
\Phi^{(\beta)}(x),\\
\label{kernel1} D^{(\beta)}_{2N}(x,y) &=&
{{\widehat{\Phi}}^{(\beta)}}(x)\prod_{N}
\Phi^{(\beta)}(y),\\
\label{kernel2} I^{(\beta)}_{2N}(x,y) &=&
-{{\widehat{\Psi}}^{(\beta)}}(x)\prod_{2N}
\Psi^{(\beta)}(y)+\delta_{1,\beta}\frac{\epsilon(x-y)}{2},
\end{eqnarray}
where
\begin{equation}
\Phi^{\bt}={(\Phi^{\bt}_0\ldots\Phi^{\bt}_n\ldots)}^t,\hspace{0.2cm}
\widehat{\Phi}^{\bt}= -{\Phi^{\bt}}^{t}Z,
\end{equation}
(similarly for $\Psi^{\bt}$) are semi-infinite vectors. They are
formed by quasi-polynomials
\begin{eqnarray}
\Phi^{(\beta)}_n(x) &=& \left(\begin{array}{c}
\phi^{(\beta)}_{2n}(x)      \\
\phi^{(\beta)}_{2n+1}(x)   \\
\end{array}\right),
\end{eqnarray}
where
\begin{eqnarray}
\label{quasipolynomial} \phi_{n}^{\bt}(x) &=&
\frac{1}{\sqrt{g^{\bt}_n}}\Pi^{\bt}_{n}(x)\exp[-V(x)],
\end{eqnarray}
and
\begin{eqnarray} \Pi^{\bt}_{n}(x) &=& \sum^{n}_{k=0}c^{(n,\beta)}_{k}x^k,
\end{eqnarray}
is the SOP of order $n$.
\begin{eqnarray}
Z &=& \left(\begin{array}{cc}
0 & 1     \\
-1 & 0         \\
\end{array}\right)
\dotplus \ldots \dotplus
\end{eqnarray}
is a semi-infinite anti-symmetric block-diagonal matrix with
$Z^2=-1$ and
\begin{eqnarray}
\nonumber \epsilon(r) &=& \frac{|r|}{r} .
\end{eqnarray}
$\delta$ is the kronecker delta. The matrix
\begin{equation}
\prod_{2N}={\rm diag}(\underbrace{1,\ldots,1}_{2N},0,\ldots, 0)
\end{equation}
have $2N$ entries. Finally, we define
\begin{equation}
\label{psi} \Psi^{(4)}_{n}(x) =
  \Phi '^{(4)}_{n}(x),\hspace{0.3cm}
\Psi^{(1)}_{n}(x)=\int\Phi^{(1)}_{n}(y)\epsilon(x-y)dy,
\end{equation}
which satisfy skew-orthonormality relation
\cite{mehta,ghosh,ghoshpandey}:
\begin{equation}
\label{ortho1} ({\tpsi}^{\bt}_n,\Phi_m^{\bt})
\equiv\int_{\Gamma}{\tpsi}^{\bt}_n{\Phi}_m^{\bt} dx=\delta_{nm}.
\end{equation}
The contour of integration `$\Gamma$' is the real axis. Here, one
must mention that under suitable assumptions on $V(x)$ that
ensures convergence of the integral (\ref{ortho1}), for a given
$V(x)$, these SOP are unique up to the addition of a lower even
order polynomial to the odd ones.

For the unitary ensemble, study of correlation function involve
similar kernel function, which is calculated \cite{ghoshpandey,
ghosh} using the well known Christoffel Darboux formula
\cite{szego}. To study the kernel functions arising in OE and SE
(\ref{kernel},\ref{kernel1},\ref{kernel2}), we derive a GCD
formula.

To this effect, we expand $x\Phi^{\bt}(x)$, ${(\Phi^{\bt}(x))}'$
and ${(x\Phi^{\bt}(x))}'$ in terms of $\Phi^{(\beta)}(x)$ (and
hence introduce the semi-infinite matrices $Q^{\bt}$, $P^{\bt}$
and $R^{\bt}$ respectively):

\begin{eqnarray}
\label{qp}
x\Phi^{\bt}(x) &=& {Q}^{\bt}\Phi^{\bt}(x),\\
\label{p} \Psi^{(4)}(x) &=& P^{(4)}\Phi^{(4)}(x),{\hspace{0.7cm}}
x\Psi^{(4)}(x)=R^{(4)}\Phi^{(4)}(x),\\
\label{r}\Phi^{(1)}(x) &=&
P^{(1)}\Psi^{(1)}(x),{\hspace{0.7cm}}x\Phi^{(1)}(x)=R^{(1)}\Psi^{(1)}(x),
\end{eqnarray}
where (\ref{r}) is obtained by multiplying the above expansion by
$\epsilon(y-x)$ and integrating by parts. They satisfy the
following commutation relations:
\begin{equation}
\label{string} [Q^{\bt},P^{\bt}]=1, \hspace{1cm}
[R^{\bt},P^{\bt}]=P^{\bt}.
\end{equation}

Using  $(\psi_{n}^{(4)}(x),\psi_{m}^{(4)}(x))$ and
$(x\psi_{n}^{(4)}(x),\psi_{m}^{(4)}(x))$ for $\beta=4$, and
replacing $\psi^{(4)}(x)$ by $\phi^{(1)}(x)$ for $\beta=1$, we get
\begin{eqnarray}
\label{dual}
P^{\bt}=-{P^{\bt}}^D,\hspace{1cm}R^{\bt}=-{R^{\bt}}^D,
\end{eqnarray}
where dual of a matrix $A$ is defined as
\begin{equation}
A^D=-ZA^{t}Z.
\end{equation}
However starting with $(x\phi_{n}^{\bt}(x),\psi_{m}^{\bt}(x))$ and
using (\ref{string}), we get
\begin{eqnarray}
Q^{\bt}={Q^{\bt}}^D+{(P^{\bt})}^{-1}.
\end{eqnarray}

We will use the  matrices $P^{\bt}$ and $R^{\bt}$ to obtain the
GCD formula. For $\beta=4$, using (\ref{kernel}), (\ref{p}) and
(\ref{dual}), we get

\begin{eqnarray}
\label{sp} \nonumber
&& S^{(4)}_{2N}(x,y)-S^{(4)}_{2N}(y,x)\\
\nonumber &=&
\left[{{\Phi}^{(4)}}^{t}(x)\prod_{2N}Z\prod_{2N}\Psi^{(4)}(y)
+{{\Psi}^{(4)}}^{t}(x)\prod_{2N}Z\prod_{2N}\Phi^{(4)}(y)\right],\\
\nonumber &=&
\left[{{\Phi}^{(4)}}^{t}(x)\prod_{2N}Z\prod_{2N}P\Phi^{(4)}(y)
+{{\Phi}^{(4)}}^{t}(x)P^t\prod_{2N}Z\prod_{2N}\Phi^{(4)}(y)\right],\\
\nonumber &=& \left[-{{\Phi}^{(4)}}^{t}(x)ZZ\prod_{2N}Z\prod_{2N}P\Phi^{(4)}(y)\right.\\
\nonumber
&& \left.+{{\Phi}^{(4)}}^{t}(x)ZZP^{t}ZZ\prod_{2N}Z\prod_{2N}\Phi^{(4)}(y)\right],\\
&=&{{\widehat{\Phi}}}^{(4)}(x)\left[P^{(4)},\prod_{2N}\right]\Phi^{(4)}(y),
\end{eqnarray}
where we have used $Z\prod_{2N}Z\prod_{2N}=-\prod_{2N}$.
Similarly, using (\ref{kernel}), (\ref{psi}), (\ref{p}) and
(\ref{dual}) we get

\begin{eqnarray} \label{sr}
\nonumber && yS^{(4)}_{2N}(x,y)- xS^{(4)}_{2N}(y,x)\\
&=& \nonumber
\left(x\frac{d}{dx}
+y\frac{d}{dy}\right){\Phi^{(4)}}^t(x)\prod_{2N}Z\prod_{2N}\Phi^{(4)}(y),\\
\nonumber &=&
\left[{{\Phi}^{(4)}}^{t}(x)R^{t}\prod_{2N}Z\prod_{2N}\Phi^{(4)}(y)
+{{\Phi}^{(4)}}^{t}(x)\prod_{2N}Z\prod_{2N}R\Phi^{(4)}(y)\right],\\
&=&
{{\widehat{\Phi}}}^{(4)}(x)\left[R^{(4)},\prod_{2N}\right]\Phi^{(4)}(y).
\end{eqnarray}
Finally using (\ref{sp}) and (\ref{sr}), the GCD for the SE
($\beta=4$) is given by
\begin{eqnarray}
\label{gcd4}
S^{(4)}_{2N}(x,y)=
\frac{x{{\widehat{\Phi}}}^{(4)}(x)\left[P^{(4)},\prod_{2N}\right]\Phi^{(4)}(y)
-{{\widehat{\Phi}}}^{(4)}(x)\left[R^{(4)},\prod_{2N}\right]\Phi^{(4)}(y)}{x-y}.
\end{eqnarray}

Following a similar procedure, the GCD for the OE ($\beta=1$) is
given by
\begin{eqnarray}
\label{gcd1} S^{(1)}_{2N}(x,y)=
\frac{y{{\widehat{\Psi}}}^{(1)}(x)\left[P^{(1)},\prod_{2N}\right]\Psi^{(1)}(y)
-{{\widehat{\Psi}}}^{(1)}(x)\left[R^{(1)},\prod_{2N}\right]\Psi^{(1)}(y)}{y-x}.
\end{eqnarray}

Here, one might recall, that for orthogonal polynomials, the
Christoffel-Darboux sum takes the form
\begin{equation}
S^{(2)}_{N}(x,y)=\phi^{t}(x)\prod_{N}\phi(y)
=\frac{\phi^{t}(x)\left[Q,\prod_{N}\right]\phi(y)}{x-y},
\end{equation}
where $Q$ is a tri-diagonal Jacobi matrix, and $\phi$ are the
normalized orthogonal quasi-polynomials. Any matrix of the form
$\left[A,\prod_{N}\right]$ has only off-diagonal blocks whose size
depends on the number of bands above and below the diagonal of
$A$. Thus the `size' of $P^{\bt}$ and $R^{\bt}$ decides the number
of terms around the `Fermi-level' that will ultimately contribute
to the correlation. For example, for the orthogonal polynomials,
the Christoffel-Darboux has only two terms.

Having derived the GCD for arbitrary weight, we will study
ensembles with
\begin{equation}
\label{V(x)} V(x)=\sum_{l=1}^{d+1}\frac{u_{l}}{l}x^{l},
\end{equation}
where $V(x)$ is a polynomial of order $d+1$. $u_{l}$ is called the
deformation parameter \cite{eynard1}. We will show that for such
ensembles, the matrix $P^{\bt}$ and $R^{\bt}$ are finite band
matrices, i.e. have finite number of bands below and above the
principal diagonal.

For $\beta=4$,
\begin{eqnarray}
\nonumber {{\phi}'}^{(4)}_{n}(x)
&=& \sum P^{(4)}_{n,m}{\phi}^{(4)}_{m}(x)\\
\nonumber
      &=&
      \left[-V^{\prime}(x)\phi^{(4)}_{n}(x)+\phi^{(4)}_{n-1}(x)+\ldots\right]\\
      &=&-\sum{(V^{\prime}(Q))}_{n,m}{{\phi}}^{(1)}_{m}(x)
      +{{\phi}}^{(1)}_{n-1}(x)+\ldots,
\end{eqnarray}
while for $\beta=1$,
\begin{eqnarray}
\nonumber {{\phi}}^{(1)}_{n}(x)
 &=& \sum P^{(1)}_{nm}{\psi}_{m}^{(1)}(x) \\
\nonumber
&=& \int\frac{d}{dy}\left[{{\phi}}^{(1)}_{n}(y)\right]\epsilon (x-y) dy\\
      &=&
\nonumber
      \int\left[-\sum{(V^{\prime}(Q))}_{n,m}{{\phi}}^{(1)}_{m}(y)\right]\epsilon(x-y)dy\\
      &&+{{\psi}}^{(1)}_{n-1}(x)+\ldots.
\end{eqnarray}
 This gives
\begin{equation}
\label{pv} \left[ P^{\bt}+V^{\prime}(Q^{\bt})\right]={\rm lower}.
\end{equation}

Similarly, for $\beta=4$, we have
\begin{eqnarray}
\label{rq4}
 \nonumber x{\psi}^{(4)}_n(x) &=& \sum
R^{(4)}_{n,m}{\phi}^{(4)}_{m}(x)
=x\frac{d}{dx}{\phi}^{(4)}_{n}(x)\\
\nonumber
      &=&
      [-xV'(x){{\phi}^{(4)}_{n}}(x)+ n{{\phi}^{(4)}_{n}}(x)+\ldots\\
      &=& -(\sum_{m,l} Q^{(4)}_{nm}[(V'(Q^{(4)}))_{ml}{{\phi}^{(4)}_{l}}(x))
      +\ldots,
\end{eqnarray}
while for $\beta=1$, we get
\begin{eqnarray}
\label{rq1} \nonumber
x{\phi}^{(1)}_{n}(x) &=& \sum R^{(1)}_{nm}{\psi}_{m}^{(1)}(x)\\
\nonumber
 &=& \int\frac{d}{dy}\left[y{\phi}^{(1)}_{n}(y)\right]\epsilon(x-y)dy\\
\nonumber
      &=&
      \int \left[-y
      V'(y){{\phi}^{(1)}_{n}}(y)\right]\epsilon(x-y)dy\\
\nonumber
      &&+(n+1){{\psi}^{(1)}_{n}}(x)+\ldots\\
      &=& -(\sum_{m,l} Q^{(1)}_{nm}[(V'(Q^{(1)}))_{ml}{{\psi}^{(1)}_{l}}(x))
      +\ldots.
\end{eqnarray}
Thus we get
\begin{equation}
\label{rv} \left[
R^{(\beta)}+Q^{(\beta)}V^{\prime}(Q^{(\beta)})\right]={\rm
lower}_{+},
\end{equation}
where  `lower' denotes a strictly lower triangular matrix and
$`{\rm lower}_{+}'$ a lower triangular matrix with the principal
diagonal. Since $Q^{\bt}$ has only one band above the diagonal,
Eqs.(\ref{pv},\ref{rv}) confirm that $P^{\bt}$ and $R^{\bt}$ has
$d$ and $d+1$ bands above and below (since they are anti-self
dual) the principal diagonal. However, unlike $\beta=2$, $Q^{\bt}$
for $\beta=1$ and $4$ is not a finite band matrix.

$Q^{\bt}$ can be calculated in terms of the normalization constant
$g^{\bt}_{n}$ and the coefficients of the polynomials
$c^{(k,\beta)}_j$. For SOP, with $g_{2N}^{\bt}=g_{2N+1}^{\bt}$, we
have
\begin{eqnarray}
\label{qo1} \nonumber Q^{\bt}_{j,j+1} &=&
\frac{c^{(j,\beta)}_{j}}{c^{(j+1,\beta)}_{j+1}}\sqrt{\frac{g^{\bt}_{j+1}}{g^{\bt}_{j}}},
Q^{\bt}_{j,j} = \frac{c^{(j,\beta)}_{j-1}}{c^{(j,\beta)}_{j}}
-\frac{c^{(j+1,\beta)}_{j}}{c^{(j+1,\beta)}_{j+1}}.
\end{eqnarray}


Finally using Eqs.(\ref{gcd4}), (\ref{gcd1}) and the asymptotic
results for these SOP \cite{ghoshpandey}, we present an
alternative derivation of the level-density and the `two-point'
function for the Gaussian orthogonal and symplectic ensembles. For
the Gaussian Ensembles, with $d=1$, (\ref{gcd4}) and (\ref{gcd1})
give

\begin{equation}
\label{s4d1} S^{(4)}_{2N}(x,y)=
\frac{[xP^{(4)}_{(1,2)}-R^{(4)}_{(1,2)}]\phi^{(4)}_{(0,2)}
+R^{(4)}_{(0,2)}\phi^{(4)}_{(1,2)}
-R^{(4)}_{(1,3)}\phi^{(4)}_{(0,3)}}{x-y},
\end{equation}
and
\begin{equation}
\label{s1d1} S^{(1)}_{2N}(x,y)=
\frac{[yP^{(1)}_{(1,2)}-R^{(1)}_{(1,2)}]\psi^{(1)}_{(0,2)}
+R^{(1)}_{(0,2)}\psi^{(1)}_{(1,2)}
-R^{(1)}_{(1,3)}\psi^{(1)}_{(0,3)}}{y-x},
\end{equation}
respectively, where
\begin{equation}
\phi^{(4)}_{(j,k)}\equiv [\phi^{(4)}_{2N+j}(x)\phi^{(4)}_{2N+k}(y)
-\phi^{(4)}_{2N+j}(y)\phi^{(4)}_{2N+k}(x)]
\end{equation}
(and similarly for $\psi^{(1)}_{(j,k)}$ ) and
\begin{equation}
A^{\bt}_{(j,k)}\equiv A^{\bt}_{2N+j,2N+k}.
\end{equation}


For $\beta=4$, using $g^{(4)}_{2n}=(2n+1)!{\pi}^{1/2}2^{2n}$ and
$c^{(n,4)}_{n}={(\sqrt{2})}^{3n-1}$ \cite{ghoshpandey}, we have
$Q^{(4)}_{(0,1)}=1/2\sqrt{2}$ and $Q^{(4)}_{(1,2)}=\sqrt{2}N$.
From Eq.(\ref{pv}) and (\ref{rv}), with $u_2=2$ and $u_1=0$, we
get
\begin{eqnarray}
P^{(4)}_{(1,2)} &=& -2Q^{(4)}_{(1,2)}=-4N/\sqrt{2},\\
R^{(4)}_{(0,2)} &=& -2Q^{(4)}_{(0,1)}Q^{(4)}_{(1,2)}=-N, \\
R^{(4)}_{(1,3)} &=& -2Q^{(4)}_{(1,2)}Q^{(4)}_{(2,3)}=-N.
\end{eqnarray}
For large $N$, we use the asymptotic results (with Gaussian
weight) for the SOP \cite{ghoshpandey}:

\begin{eqnarray}
\phi^{(4)}_{2n+1}(x) &=&
\frac{\sin\left[f^{(4)}(n,\theta)\right]}{n^{1/4}\sqrt{\pi\sin\theta}},\\
 \phi^{(4)}_{2n}(x) &=& \frac{1}{{(4n)}^{1/4}}
\left[\frac{\cos\left[f^{(4)}(n,\theta)\right]}{2\sqrt{2n\pi{\sin}^3\theta}}
                  +\frac{1}{2}\right],
\end{eqnarray}
where
\begin{eqnarray}
\nonumber
f^{(4)}(n,\theta) &=& (n+3/4)(\sin 2\theta-2\theta)+\frac{3\pi}{4},\\
            &=& 2\int_{-\sqrt{2n}}^x \rho(n,x)dx+\frac{3\pi}{4},
\end{eqnarray}
and $x={(2n+3/2)}^{1/2}\cos\theta $. For a given $x$, $\theta$
depends on $n$; for example ${\theta}_{n}-{\theta}_{n\mp
1}\simeq\pm {(2n\tan\theta_{n})}^{-1}$. Then with $y=x+\Delta x$
and $\theta\equiv{\theta}_{n}$, where
${\theta}_{n}\equiv{({\theta}_{2n}{\theta}_{2n+1})}^t $, and
expanding in $\theta$ and $N$, we get from the first term in
Eq.(\ref{s4d1}):
\begin{equation}
xP^{(4)}_{(1,2)}[\phi^{(4)}_{(0,2)}] =
-\frac{1}{4\pi}\frac{\cos\theta}{\sin^3}
\sin\left(\Delta\theta\frac{\partial
f^{(4)}(N,\theta)}{\partial\theta}\right) \sin 2\theta.
\end{equation}
The second and third terms in Eq.(\ref{s4d1}) give
\begin{eqnarray}
\nonumber R^{(4)}_{(0,2)}[\phi^{(4)}_{(1,2)}]
&=&-R^{(4)}_{(1,3)}[\phi^{(4)}_{(0,3)}]
\\
&=& \frac{\cos 2\theta}{4\pi\sin^2\theta}
\sin\left(\Delta\theta\frac{\partial
f^{(4)}(N,\theta)}{\partial\theta}\right).
\end{eqnarray}
We know that the odd SOP is arbitrary to the addition of a lower
even order polynomial. This choice cancels the term
$R^{(4)}_{(1,2)}[\phi^{(4)}_{(0,2)}]$. Collecting all the terms,
we get
\begin{eqnarray}
{S^{(4)}_{2N}(x,y)}=\frac{\sin[2\sqrt{(2N-x^2)}\Delta
x]}{2\pi\Delta x},\hspace{0.5cm}|x|<\sqrt{2N}.
\end{eqnarray}
Taking $\Delta x\rightarrow 0$, we get the famous `semi-circle',
while
\begin{eqnarray}
\frac{S^{(4)}_{2N}(x,y)}{S^{(4)}_{2N}(x,x)}=\frac{\sin[2\pi\Delta
xS^{(4)}_N(x,x)]}{2\pi\Delta x S^{(4)}_{2N}(x,x)}=\frac{\sin 2\pi
r}{2\pi r},
\end{eqnarray}
where $r=\Delta x S^{(4)}_{2N}(x,x)$, gives the universal
sine-kernel in the bulk of the spectrum.

For $\beta=1$, we have from \cite{ghoshpandey}
$g^{(1)}_{2n}=(2n)!{\pi}^{1/2}2^{2n}$ and
$c^{(2n,1)}_{2n}=-c^{(2n+1,1)}_{2n+1}={2}^{2n}$, which gives
$Q^{(1)}_{(0,1)}=-1$, $Q^{(1)}_{(1,2)}=-N$. For $u_2=1$ and
$u_1=0$, Eqs.(\ref{pv}) and (\ref{rv}) gives
\begin{eqnarray}
&& P^{(1)}_{(1,2)} = -Q^{(1)}_{(1,2)}=N,\\
&& R^{(1)}_{(0,2)} = -Q^{(1)}_{(0,1)}Q^{(1)}_{(1,2)}=-N,\\
&& R^{(1)}_{(1,3)} = -Q^{(1)}_{(1,2)}Q^{(1)}_{(2,3)}=-N.
\end{eqnarray}
For large $N$, we use the asymptotic results for the SOP
\cite{ghoshpandey}:
\begin{eqnarray}
\psi^{(1)}_{2n+1}(x) &=&
\frac{\sin\left[f^{(1)}(n,\theta)\right]}{n^{1/4}\sqrt{\pi\sin\theta}},\\
\psi^{(1)}_{2n}(x) &=& -\frac{1}{2{n}^{1/4}}
\left[\frac{\cos\left[f^{(1)}(n,\theta)\right]}
{\sqrt{n\pi{\sin}^3\theta}}\right],
\end{eqnarray}
where
\begin{eqnarray}
\nonumber
f^{(1)}(n,\theta) &=& (n+1/4)(\sin 2\theta-2\theta)+\frac{3\pi}{4},\\
            &=& \int_{-\sqrt{4n}}^x \rho(n,x)dx+\frac{3\pi}{4},
\end{eqnarray}
and $x={(4n+1)}^{1/2}\cos\theta$.
 Writing $y=x+\Delta x$, and expanding in $\theta$ and $N$
 in (\ref{s1d1}), we get
\begin{eqnarray}
{S^{(1)}_{2N}(x,y)}=\frac{\sin[\sqrt{(4N-x^2)}\Delta x]}{\pi\Delta
x},\hspace{0.5cm}|x|<\sqrt{4N}.
\end{eqnarray}
Taking $\Delta x\rightarrow 0$, we get the level density, while
\begin{eqnarray}
\frac{S^{(1)}_{2N}(x,y)}{S^{(1)}_{2N}(x,x)}=\frac{\sin[\pi
S^{(1)}_{2N}(x,x)\Delta x]}{\pi\Delta x
S^{(1)}_{2N}(x,x)}=\frac{\sin\pi r}{\pi r},
\end{eqnarray}
where $r=\Delta x S^{(1)}_{2N}(x,x)$, gives the universal
sine-kernel in the bulk of the spectrum.

For general $d$ the correlation function corresponding to a weight
with single support can be obtained using the asymptotic results
for the SOP \cite{eynard}. However, one needs to understand in
greater detail the structure of the finite-band matrices and hence
the matrix $Q^{(\beta)}$ to come up with a proof.

In conclusion, the unitary ensembles of random matrices, which
involve the orthogonal polynomials have been well studied in
recent years. In contrast, barring a few specific weights, nothing
much is known about the OE and SE. This is mainly due to the
hurdles created by the SOP.

In this paper, we have made some progress in understanding some of
the basic properties of these SOP. In this context, we would like
to emphasize that the GCD formula, derived in this paper, can at
best be considered as the first step for a systematic study of the
OE and SE of random matrices. One still needs to develop the
theory further to come to an equal footing with the unitary
ensemble of random matrices. For example, one would like to
understand in greater details the asymptotic behavior of these SOP
\cite{eynard}
 to study different correlations for a larger family of OE and
SE. We would also like to point out the similarity in the GCD
formula for $\beta=1$ and $4$ with the interchange of $\Phi$ with
$\Psi$. This may be useful in proving the duality between these
two ensembles. We believe that these SOP satisfy a $d\times d$
differential system, which can be used to formulate a
Riemann-Hilbert problem for these matrix models. We wish to come
back to a few of these questions in a later publication.

\vspace{0.2cm} I am grateful to Bertrand Eynard for useful
discussions.

\end{document}